# Identifying the phase diagram structure for optimal information integration in morphogen systems

Kakit Kong[1], Chunxiong Luo[1,2,3], Feng Liu[2,4*]

[1] *The State Key Laboratory for Artificial Microstructures and Mesoscopic Physics, School of Physics, Peking University, Beijing, 100871, China*

[2] *Center for Quantitative Biology, Peking University, Beijing, 100871, China*

[3] *Wenzhou Institute University of Chinese Academy of Sciences, Wenzhou, Zhejiang, China*

[4] *Key Laboratory of Hebei Province for Molecular Biophysics, Institute of Biophysics, School of Health Science & Biomedical Engineering, Hebei University of Technology, Tianjin, 300130, China*

## Abstract

Gene regulatory networks (GRNs) perform a wide range of biological functions. It is, however, often challenging to reveal their functioning mechanism with the conventional approach focusing on the network topological structure from a bottom-up perspective. Here, we apply the top-down approach based on the optimality theory to study the information integration in morphogen systems, and show that the optimal integration strategy raises requirement on the phase diagram, rather than the topological structure, of a GRN. For the morphogen system in early fly embryos, our parameter-free model can quantitatively predict the patterning position shifts upon the dosage change of the morphogen Bicoid.

An essential task in physical biology is to reveal the functioning mechanism of GRNs [1]. To achieve this goal, many studies describe a GRN as a topological graph emphasizing its gene interaction at the molecular level, and model it from a bottom-up perspective. Usually, a set of differential equations is established to formulate the time evolution of gene expression. Since many parameters in these models cannot be determined experimentally, on the one hand, the "data-driven" scheme has been widely used to fit the parameters such that the model can reproduce the observed gene expression dynamics [2]. On the other hand, the "function-oriented" approach has been applied to search for robust structural modules by screening the parameter space [3].

Although the bottom-up modelling can capture details of GRNs and provide insights on their functions, they are either intensively data-dependent or computationally expensive. A complementary approach to reveal the functioning mechanism of GRNs would be top-down modelling based on fundamental principles, such as the optimality theory [4,5]. For example, by optimizing the direction inference against noisy sensory signals, the resulting model of a chemotactic system based on "information-theoretically optimal dynamics" is equivalent to the experiment-based biochemical model [6]. This study implies the possibility of using optimality theory as the foundation of modeling. Moreover, optimality theory can constrain the size of parameter space in mathematical models with high degree of freedom, such as non-equilibrium models [5]. However, it has yet to demonstrate that the functioning mechanism of a GRN can be revealed with the top-down modelling.

Morphogen systems, as an example of information integration, are ideal systems to study from a top-down perspective using the optimality theory. As diffusive transcription factors distributing over the organism, morphogens provide positional cues for cells to differentiate to their respective fates [7]. However, since the morphogen distribution inevitably contains noise, accurate and precise differentiation necessitates a non-trivial integration of information from multiple morphogens [8–10]. For example, the zygotic gene expression pattern in early fly embryos is highly precise and scaling despite the more noisy and non-scaling maternal morphogens [11–13]. The gap gene circuit model suggests that the cross-regulation of the gap gene network is important in noise filtering and scaling [14,15]. While the bi-gradient model proposed that the high precision and scaling of the gap gene patterning in the middle of the embryo result from the noise cancelation of two correlated morphogen gradients including the Bicoid (Bcd) gradient originated from the anterior pole, and a gradient X originated from the posterior pole with the same shape as the Bcd gradient [10]. However, the gradient X has yet to be discovered. Recently, the precision of the embryonic scale gene expression pattern has been shown to reach the theoretical limit [16]. Therefore, an appealing postulation is that morphogen systems, such as the patterning network in fly embryos, might be optimized against noise to guarantee a reliable development [17].

In this letter, we study the optimal information integration strategy in morphogen systems from the top-down perspective. Comparing to the work showing that an optimal decoder can predict downstream gene expression pattern [17], we aim to reveal how optimality is realized by a GRN. Our parameter-free model identifies a characteristic phase diagram structure for the optimal information integration. By applying this model to the patterning network in early fly embryos, we can quantitatively predict the response of different patterns to the morphogen dosage adjustment, implying that optimal information integration is the mechanism for the high patterning precision.

*The phase diagram structure representing information integration strategy in morphogen systems.* To generalize the bi-gradient model [10], we consider two unscaled 1D morphogens [Fig. 1(a)], $M_1$ and $M_2$, and assume their concentration distribution depends on the absolute distance from the anterior pole and the posterior pole, respectively, since any morphogen without spatial-regulated diffusion or degradation is unscaled with the system size [18]. And the two morphogens do not need to be the same shape [10]. Let $m_1$ and $m_2$ denote the concentration of morphogen $M_1$ and $M_2$, respectively, $\langle m_1 \rangle = f_1(xL)$ and $\langle m_2 \rangle = f_1[(x-1)L]$, where $\langle ... \rangle$ denote the ensemble average, $f_1$ and $f_2$ denote two different functions, $x$ is the relative position ranging from 0 to 1, and $L$ is the system size normalized by the average size.

We consider only the extrinsic noise in our model since intrinsic noise can be attenuated by spatiotemporal average [7,19,20]. Specifically, each morphogen contains two extrinsic noise components: an uncorrelated Gaussian extrinsic noise, with the position dependent amplitude of $\delta_1 = \delta_1(x)$ and $\delta_2 = \delta_2(x)$ for $M_1$ and $M_2$, respectively; and the correlated extrinsic noise originated from $L$ fluctuation, which is assumed to be Gaussian with the standard deviation of $\delta_L$. Therefore, the spatial concentration distribution of the two morphogens follows:

$$\begin{array}{rl} m_1(x) = & f_1(xL) + \delta_1 \gamma_1 \\ m_2(x) = & f_2[(x-1)L] + \delta_2 \gamma_2 \end{array}, \quad (1)$$

where $L = 1 + \delta_L \gamma_L$. $\gamma_1, \gamma_2$ and $\gamma_L$ are independent Gaussian random variables with zero mean and unit variance. The average distribution of the two morphogens can be plotted on the $m_1 - m_2$ diagram as a parametric equation, i.e., the "input curve" in Fig. 1(b) [see supplemental material (SM) [21] and Fig. S1 for details].

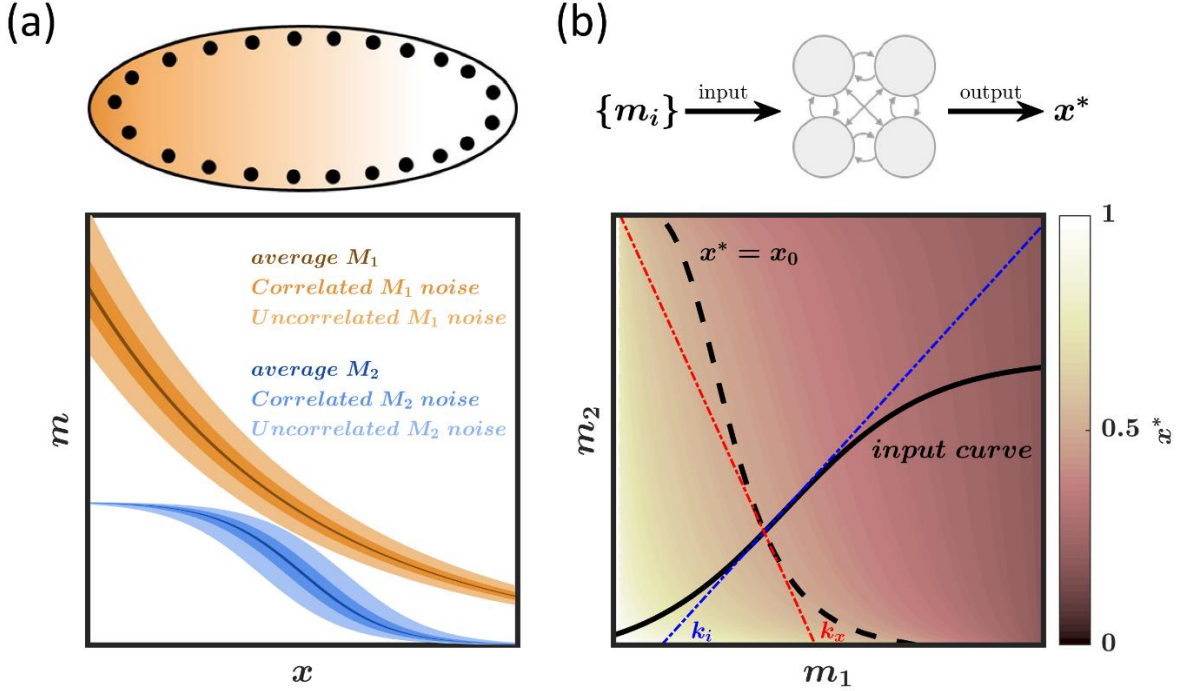

FIG. 1. Positional information readout problems. (a) In a multicellular organism such as a fly embryo, two morphogens distribute over the organism with both correlated and uncorrelated extrinsic noise. (b) A coarse-grained GRN reads the local concentration of two morphogen gradients encoding positional information, and outputs an inferred position $x^*$ for a given nucleus. The input-output relation [Eq. (2)] can be represented by a phase diagram. Different colors represent different values of the inferred position. The black curve is the input curve [Eq. (1)]; the black dashed line is an example of contours $x^* = x_0$. Red and blue dotted line show the tangent of a contour of the inference function and the input curve, respectively. And $k_x$ and $k_i$ represent their corresponding slopes.

Generally, the response of cells to the morphogen signals can be described by a vector variable, e.g., the expression level of a set of genes. We first assume their response can be parameterized by a single variable for simplicity. We discuss the general case in SM [21]. Without losing generality, we use the "inferred position" ($x^*$) to parameterize the behavior of a cell. "Inferred" means $\langle x^* \rangle = x$. Ideally, $x^*$ is the output of the GRN reading the input morphogen concentrations, e.g., the local $m_1$ and $m_2$ [Figs. 1(a) and 1(b)]. By coarse-graining the GRN, $x^*$ follows an inference function:

$$x^* = F(m_1, m_2). \qquad (2)$$

We further define different $x^*$ as different cell state, therefore, the $m_1 - m_2$ diagram resembles a phase diagram. At the intersection point of the input curve [Eq. (1)] and the contour of $x^*$ [Eq. (2)] on the $m_1 - m_2$ diagram, the slope of the input curve is $k_i = [f_2'(x-1)/f_1'(x)]$, and the slope of the contour is $k_x = -\left[\frac{\partial F(m_1,m_2)/\partial m_1}{\partial F(m_1,m_2)/\partial m_2}\right]_{[f_1(x),f_2(x-1)]}$ [Fig. 1(b)].

As the morphogen concentration fluctuates due to noise, the inferred position also fluctuates ("inference noise"). Or, conversely, the position $(x)$ where a given behavior $(x^*)$ occurs has a "positional noise" of (SM [21]):

$$\sigma_{x|x^*}^2 = \frac{k_x^2\sigma_1^2 + k_i^2\sigma_2^2 + \delta_L^2(k_x x^* - k_i x^* + k_i)^2}{(k_i - k_x)^2}, \quad (3)$$

where $\sigma_1 = \sigma_1(x) \triangleq |\delta_1(x)/f_1'(x)|$ and $\sigma_2 = \sigma_2(x) \triangleq |\delta_2(x)/f_2'(x-1)|$ stand for the extrinsic positional noise of the two morphogens.

This calculation unveils that, by integrating information from two morphogen sources, the output noise is controlled by two parameters: $k_x$ and $k_i$, which are two characteristic slopes on the phase diagram [Fig. 1(b)]. That is, by representing different input states as different points on the phase diagram, the information integration strategy can be describe by the geometry of the phase diagram.

***The optimal condition for information integration.*** Next, we derive the optimal condition and a relaxed condition for controlling the output noise.

Since for given spatial morphogen distribution, $k_i$ is fixed, minimizing the output noise $\sigma_{x|x^*}^2$ requires $d\sigma_{x|x^*}^2/dk_x = 0$ and $d^2\sigma_{x|x^*}^2/dk_x^2 > 0$, which yields the optimal condition: $k_x/k_i = R_{opt}$, where $R_{opt}$ is:

$$R_{opt} = -\frac{\sigma_2^2 + (1-x^*)\delta_L^2}{\sigma_1^2 + x^*\delta_L^2}. \quad (4)$$

The minimal output noise follows:

$$\sigma_{min}^2 = \frac{\sigma_1^2\sigma_2^2 + \delta_L^2[(1-x^*)^2\sigma_1^2 + x^{*2}\sigma_2^2]}{\sigma_1^2 + \sigma_2^2 + \delta_L^2}. \quad (5)$$

Another useful but relaxed condition is derived by optimizing against the **S**ize-fluctuation-**I**nduced (SI) noise only (i.e., $\delta_1 = \delta_2 = 0$), which reflects the pure scaling feature. Under this condition, combining Eq. (1) to trace out $L$ (note that $\langle x^* \rangle = x$ being the definition of inferred position) yields the SI condition that the inference function should follow $x^* = F_{SI}(m_1, m_2)$, where $F_{SI}(m_1, m_2)$ is:

$$F_{SI}(m_1, m_2) = \frac{f_1^{-1}(m_1)}{f_1^{-1}(m_1) - f_2^{-1}(m_2)}. \tag{6}$$

Under the SI condition, the output noise is:

$$\sigma_{SI}^2 = (1 - x^*)^2 \sigma_1^2 + x^{*2} \sigma_2^2. \tag{7}$$

***The equivalence between the optimal condition and the SI condition.*** We first focus on the central region of the system as in the classical bi-gradient model [10]. We find that the two conditions [Eqs. (5) and (7)] yield the same output noise at the middle ($x^* = 0.5$), assuming two morphogens have the same noise level, as

$$\sigma_{out}^2 = \sigma_1^2/2 = \sigma_2^2/2, \tag{8}$$

given $\sigma_1^2 = \sigma_2^2$. This result implies that the two conditions are similar.

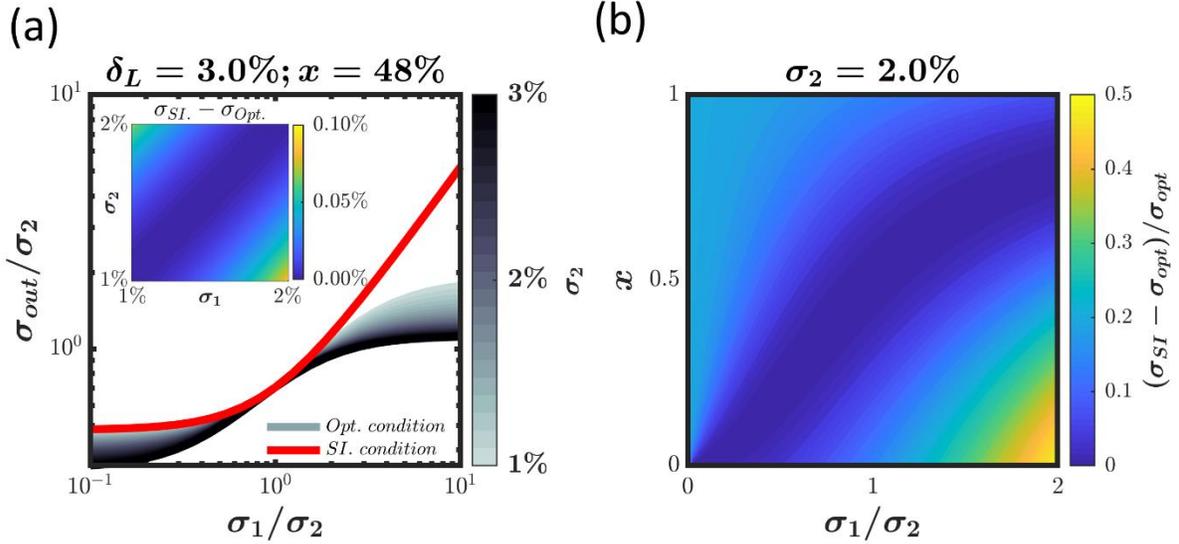

FIG. 2. The equivalence between the optimal condition and SI condition. (a) Using the position $x = 0.48$ as an example, the optimal output noise varies as $\sigma_2$ varies (between 1-3%). And the two conditions yield the same output noise given that the two input positional noises are of the same magnitude. Inset shows the difference of the output noise under SI and optimal condition as a function of the input positional noises of the two morphogen gradients. (b) The relative difference of the output noises under the two conditions is very small at nearly all positions when the noises of the two morphogens are at the same magnitude, here we use $\sigma_2 = 2\%$ as an example.

We further test this similarity by letting $\sigma_1^2 \neq \sigma_2^2$ and $x^* \neq 0.5$. We find that when the two morphogen noises are of the same magnitude, the output noises under the two condition are similar, especially in the central region, e.g., when $\sigma_1^2 = \sigma_2^2$, the relative difference between output noises under the two condition is less than 10% in $0.21 \leq x \leq 0.79$; when $\sigma_1 = \frac{1}{2}\sigma_2$ or $\sigma_1 = 2\sigma_2$, the relative difference is less than 10% in $0 \leq x \leq 0.4$ and $0.63 \leq x \leq 0.97$, respectively [Figs. 2(a) and 2(b)]. Therefore, we use the SI condition, instead of the optimal condition, to test the noise control mechanism in fly embryos since it provides a parameter-free criterion for validation.

***The morphogen system in fly embryos.*** We apply the optimal information integration framework to study the patterning in fly embryos.

In early fly embryogenesis, Bcd is one of the most intensively studied maternal morphogens [11,22–25]. Its steady-state average concentration has an exponential distribution along the anterior-posterior axis:

$$\langle b \rangle = D f_1(xL) = D \exp(-xL/\lambda), \tag{9}$$

where $b$ denotes the Bcd concentration, $D$ is an experimentally tunable parameter representing the relative dosage of Bcd ($D = 1$ for the wildtype), and $\lambda = 0.165$ is its normalized length constant based on the experiment [26]. The embryo length $L$ shows 3% natural fluctuation [18,27]. Hunchback (Hb), being an extensively studied downstream gene, is a characteristic output of the morphogen system. Its anterior expression is regulated by Bcd, and shows a step like profile whose boundary positioned at $x_{Hb} = 0.48$ with 1% noise [12,27]. Hence, the precision of $x_{Hb}$ is later used as an example to validate our model. First, we focus on the Bcd dosage.

***The dosage response in fly embryos.*** Assuming the patterning system in fly embryos operates under the SI condition and choosing Bcd as $M_1$ in our model, combining Eqs. (1), (6) and (9) yields an average dosage response function (SM [21]):

$$\langle x \rangle = x^* + \lambda(1 - x^*) \ln D. \tag{10}$$

This equation, quantitatively predicting the position shift of a pattern denoted by $x^*$ (e.g., $x^* = 0.48$ for $x_{Hb}$) in response to the Bcd dosage adjustment (i.e., $D \neq 1$), is used as the criterion to validate the noise control mechanism.

We find that our prediction contradicts that of classical threshold dependent model, which is $\langle x \rangle = x^* + \lambda \ln D$ [Fig. S3(b), SM [21]], but aligns well with the experimental data [26] (Fig.3). This work [26] adjusted Bcd dosage and quantified the position shift of its downstream patterning including the boundaries of Hb and Kruppel (Kr), the Even-skipped (Eve) peaks, and the cephalic furrow (CF) (Fig. S2, SM [21]). In our model, different patterns correspond to different $x^*$ values which equal to their average positions in the wildtype. Most of the existed gene network models cannot quantitatively predict these patterning position deviations except the bi-gradient model as far as we know (e.g., [28]). However, the classical bi-gradient model can only be applied to patterns in the middle of embryos, while our model can be applied to patterns at all positions. Moreover, our model uses one single dosage response function, i.e., a master curve [29], to explain the behavior of all measured patterns without any parameter fitting. Since these different patterns are regulated by different sets of transcription factors with different detailed GRN structures [22], our results suggest that despite the difference in the topological structure, these GRNs could all adopt the same local phase diagram structure.

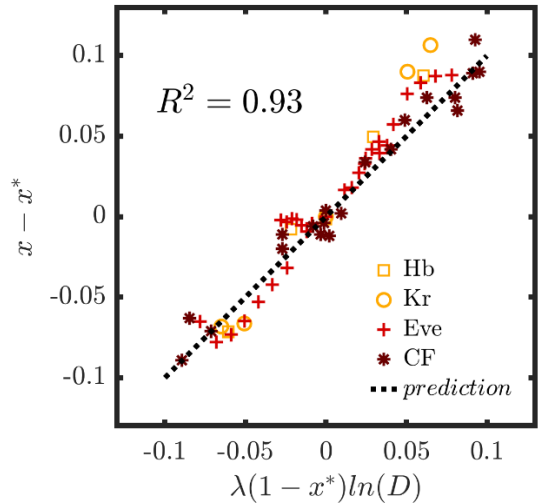

FIG. 3. Prediction of the response of the gap gene network to the Bcd dosage variation is confirmed with the experimental data [26] (Fig. S2). Assuming Bcd is morphogen $M_1$ in our model, and the GRN achieves the optimal positional information transmission under the SI condition, the position where a specific fate ($x^*$) occurs should obey the Eq. (10) when adjusting the Bcd dosage. The position shifts of the Hb boundary, Kr boundaries, Eve peaks and CF upon Bcd dosage variation aligns well with our prediction ($R^2 = 0.93$ after linearization).

Moreover, we confirm that the dosage response prediction based on the optimal condition described by Eq. (4) is nearly the same as that based on the SI condition [Fig. S3(a), SM [21]], suggesting that the patterning network in fly embryos operates under the optimal information integration condition to minimize the output noise.

***The input-output noise level.*** Based on the identified phase diagram structure, we build a simplified simulation framework to numerically investigate the required Bcd noise level for the 1% output noise of $x_{Hb}$. We assume $M_2$ has the same positional noise level as Bcd (i.e., $\sigma_1^2 = \sigma_2^2$). We find that if we incorporate the intrinsic noise in Bcd, the calculated noise level of Bcd [Fig. 4] is consistent with the experimental value [27] (SM [21]). Moreover, our results are consistent with a previous study indicating that the exponential profile can generate maximally precise information when the extrinsic noise and intrinsic noise are at the similar level [Fig. 4(b)] [30].

Our model makes no assumption on what $M_2$ is. However, the extrinsic positional noise of $M_2$ should be less than 1.4% assuming $\sigma_1^2 \approx \sigma_2^2$ [Eq. (8)]. Our experiment using LlamaTag [31,32] indicates that the maternal Hb (mHb, another well-known maternal morphogen [33]) does not satisfy the requirement of $M_2$ as its observed positional noise is $\tilde{\sigma}_{mHb} \approx 4.3\%$ (Figs. S4 and S5, SM [21]). To answer what $M_2$ is requires further studies, and the prediction of our model can be a guide toward this question.

Our study shows the great potential of the "top-down" approach in revealing the functioning mechanism of a GRN. While the "bottom-up" approach may accurately reproduce many phenomena, the potential overfitting might hinder our understanding toward the functioning mechanism. For example, although previous models for the gap gene network fit the average gene expression profiles in the wildtype, it is still unsure that they capture the core function of information integration as their prediction on the mutants' profiles are rather limited [2,34]. In contrast, a recent gap gene network model replicates many mutant phenotypes by following the scaling principle [35]. Without any parameter fitting, our model quantitatively predicts that the patterning position shifts upon Bcd dosage perturbation as a consequence of optimality.

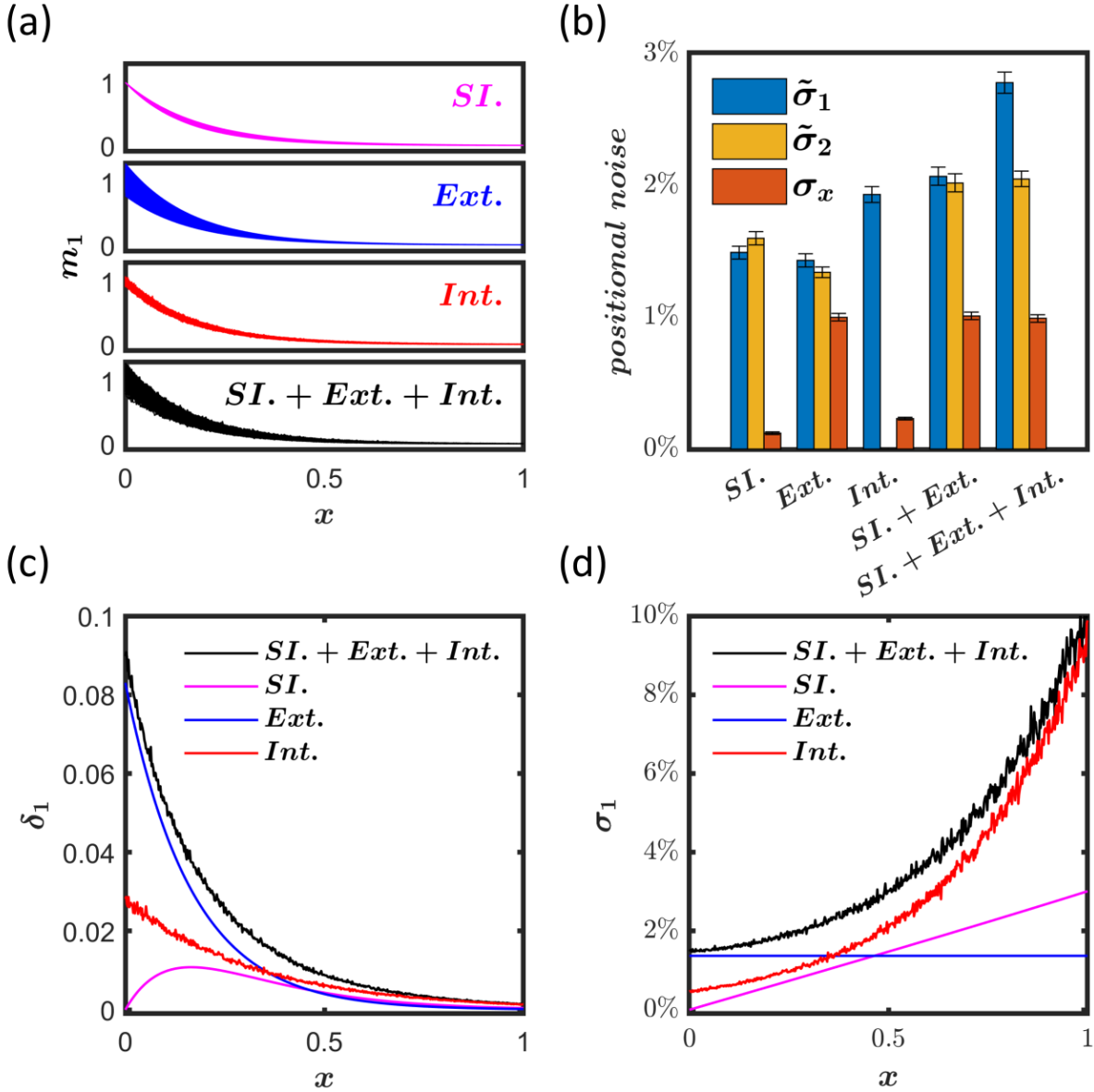

FIG. 4. Contribution of different noise components of $M_1$ to the output positional noise based on simulation. (a) The normalized concentration distribution of $M_1$ containing different noises including the correlated extrinsic noise originated from the embryo length variability (SI.), the uncorrelated extrinsic noise originated from gradient amplitude fluctuation (Ext.), and the intrinsic noise (Int.). (b) The simulated observed positional noise ($\tilde{\sigma}$) of the two morphogen gradients and output positional noise under different conditions. Intrinsic noise was only assigned to $M_1$. (c-d) The intensity standard deviation (c) and positional noise (d) of $M_1$ as a function of $x$. In simulations, the Ext noise $\sigma_1$ is assigned as a uniform value. Since $\langle m_1 \rangle = f_1(xL)$, the positional noise contributed from SI noise linearly increases toward the posterior. Int noise is assumed to be Poissonian therefore its intensity variance is proportional to the average intensity.

The optimality theory is a powerful tool in the top-down modeling. Biological systems are well known to function optimally, reaching the physical limit through evolution [4]. Among optimality, one emerging topic is the property of optimal information transmission [17]. Some studies indicate that maximum-likelihood estimation is the best decoder and demonstrate the optimal encoding strategy for multiple morphogens [8,36,37]. And in fly embryos, the pair-rule gene network seems to follow the optimal decoding according to the Bayesian inference [17]. However, since different systems have different to-be-optimized features, we would expect different target functions should be properly chosen to apply the optimality theory in the top-down modeling.

Although structural modules of GRNs are extensively studied [38–42], the connection between network structure and function is loose, e.g., the function of a GRN is usually context-dependent and a given function can be achieved by multiple network structures [43,44], while the dynamical module capturing specific behavior of systems has a tighter connection with function [45,46]. In fact, the phase diagram structure in our model is a special case of dynamical module (considering only the asymptotic behavior of a dynamic system). Therefore, to achieve optimal information integration, the phase diagram structure could be a more informative lens than the topological structure in characterizing the GRN architecture. Comparing with the traditional scheme constructing the topological structure of a GRN before calculating the phase diagram for analysis, we propose a reversed scheme that a phase diagram can be determined first, then used as a constraint for reconstructing the topological structure.

This project is supported by the National Natural Science Foundation of China 31670852 and 11974002. The modeling optimization was performed on the High Performance Computing Platform of the Center for Life Sciences, Peking University.

*liufeng-phy@pku.edu.cn